\documentclass{PoS}
\usepackage{braket}
\usepackage{amsmath}
\usepackage{amsbsy}
\usepackage{amsfonts}
\usepackage{amssymb}
\usepackage{graphics}
\usepackage{graphicx}
\usepackage{capt-of}
\usepackage{csquotes}

\title{First measurements of the $\phi_3$-sensitive decay $B^{\pm} \to D(K_{\rm S}^0 \pi^+ \pi^- \pi^0) K^{\pm}$ with Belle}

\ShortTitle{$\phi_3$ measurements from $B^{\pm} \to D(K_{\rm S}^0 \pi^+ \pi^- \pi^0) K^{\pm}$ decays}

\author{\speaker{P. K. Resmi}%
        \thanks{on behalf of the Belle Collaboration}\\
       Indian Institute of Technology Madras\\
       E-mail: \email{resmipk@physics.iitm.ac.in}}

\abstract{ The current experimental uncertainty on the CKM unitarity triangle angle $\phi_3$ is significantly larger than that on the standard model prediction. A more precise measurement of $\phi_3$ is crucial for testing the SM description of $CP$ violation and probing for new physics effects. The precision can be improved by exploring new $B$ and $D$ decay modes. The first model-independent measurement of the CKM angle $\phi_3$ using $B^{\pm} \to D(K_{\rm S}^0 \pi^+ \pi^- \pi^0) K^{\pm}$ decays is presented here. The GGSZ method is used by binning the five-dimensional phase space of the $D$ decay. This analysis uses the measurement of the average strong-phase differences across the phase space in $D\to K_{\rm S}^0 \pi^+ \pi^- \pi^0$ decays from CLEO-c, as input. The results are obtained from the full Belle data set with an integrated luminosity of 711~fb$^{-1}$ collected at the $\Upsilon(4S)$ resonance.
}

\FullConference{ European Physical Society Conference on High Energy Physics - EPS-HEP2019 -\\
			10-17 July, 2019\\
			Ghent, Belgium}

\begin{document}

\section{Introduction}

The current best measurement of the Cabibbo-Kobayashi-Maskawa~\cite{CKM} unitarity triangle angle $\phi_3$, combining all the results from different experiments, is  ($73.5^{+4.2}_{-5.1}$)$^{\circ}$~\cite{HFLAV}. This large uncertainty compared to $\phi_1$ \cite{HFLAV} is due to the small branching fractions of the decays sensitive to $\phi_3$. The value of $\phi_3$ estimated indirectly from other parameters of the unitarity triangle is ($65.3^{+1.0}_{-2.5}$)$^{\circ}$~\cite{HFLAV}. Any discrepancy between these results would imply that there are new physics effects beyond the standard model~(SM). The associated uncertainties have to be comparable for a meaningful comparison. So, improving the precision on $\phi_3$  measurement is crucial to test the $CP$ violation mechanism in the SM.

The angle $\phi_3$ is measured from the interference between the amplitudes of the color-favored $B^-\to D^0 K^-$ decay and color-suppressed $B^-\to \overline{D^0} K^-$ decay. The corresponding amplitudes can be written as $A_{\rm fav} = A$ and $A_{\rm sup} = Ar_{B}e^{i(\delta_{B} - \phi_{3})}$, where $\delta_{B}$ is the strong phase difference between the decay processes, and 
\begin{equation}
r_{B} = \frac{\mid A_{\rm{sup}} \mid} {\mid A_{\rm{fav}}\mid}.
\end{equation}
The statistical uncertainty on $\phi_3$ scales as $1/r_B$. The value of $r_B$ is approximately equal to 0.1 for $B^{\pm}\to DK^{\pm}$ decays and 0.005 for $B^{\pm}\to D\pi^{\pm}$ decays, which means the latter decay is not very sensitive to $\phi_3$. However, $B^{\pm}\to D\pi^{\pm}$ decays serve as an excellent calibration sample due to the similar topology as $B^{\pm}\to DK^{\pm}$ decays and larger sample size. The simultaneous analysis of these two decay modes allows the extraction of the cross-feed background from the misidentification of a pion as a kaon, and {\it vice versa}, from data. The results presented here are a summary of the analysis given in Ref.~\cite{Resmi-Belle}

\section{Formalism}

There are different methods to determine $\phi_3$ depending on the $D$ final state of interest. Here, we study the four-body self-conjugate state $D \to K_{\rm S}^0\pi^+\pi^-\pi^0$, which has a branching fraction of 5.2\%~\cite{PDG}. The intermediate resonances like $K_{\rm S}^0\omega$ and $K^{*}\rho$ allow for a model-independent determination of $\phi_3$ from this single channel by analysing the $D$ phase space regions~\cite{GGSZ,GGSZ2}. In this method, the $CP$ violation sensitive parameters are measured in independent regions of $D$ phase space called \enquote{bins}.

The decay rates of $B^{\pm} \to DK^{\pm}$ candidates in each bin is given as
\begin{equation}
\Gamma_{i}^{\pm} \propto K_{i} + r_{B}^{2}\overline{K_{i}} +2\sqrt{K_{i}\overline{K_{i}}}(c_{i}x_{\pm} \mp s_{i}y_{\pm}), \label{Eq:GGSZ}
\end{equation} 
\noindent where $x_{\pm} = r_{B}\cos (\delta_{B} \pm \phi_{3})$ and $y_{\pm} = r_{B}\sin (\delta_{B} \pm \phi_{3})$. Here, $K_{i}$ and $\overline{K_{i}}$ are the fraction of flavour-tagged $D^{0}$ and $\overline{D^{0}}$ events in the $i^{\rm th}$ bin, respectively. The parameters $c_{i}$ and $s_{i}$ are the amplitude-weighted average of the cosine and sine of the strong-phase difference between $D^{0}$ and $\overline{D^{0}}$ over the $i^{\rm th}$ bin. The $\phi_3$-sensitive parameters $x_{\pm}$ and $y_{\pm}$ can be determined with the total number of bins $\mathcal{N}\geq$3, provided the parameters $K_i$, $\overline{K_i}$, $c_i$ and $s_i$ are measured from elsewhere. The values of $K_{i}$ and $\overline{K_{i}}$ are determined from from $D^{*+} \to D^0 \pi^{+}$ decays with good precision due to the large sample size available. The $c_i$ and $s_i$ parameters are estimated from CLEO-c, where quantum-entangled $D^{0}\overline{D^{0}}$ pairs are produced via $e^{+}e^{-} \to \psi(3770) \to D^{0}\overline{D^{0}}$.

\section{$c_i$ and $s_i$ measurements from CLEO-c}

The $c_i$ and $s_i$ parameters are measured from a data sample corresponding to an integrated luminosity of 0.8~fb$^{-1}$ collected by the CLEO-c detector.  The $D$ phase space is divided into nine independent bins around different intermediate resonances. The binning is implemented sequentially. The rates of decays of the quantum-correlated $D^{0}\overline{D^{0}}$ pairs are proportional to the $c_i$ and $s_i$ parameters. So the measurements of the decay rates of events, in which one $D$ meson is reconstructed in the signal mode $K_{\rm S}^0\pi^+\pi^-\pi^0$ and the other $D$ meson in a variety of tag modes, give sensitivity to $c_i$ and $s_i$ parameters. The results~\cite{Resmi} are given in Table~\ref{Table:cisi}.

\begin{table} [ht!] 
\centering
\begin{tabular} {l c c  c c} 
\hline 
Bin no. & Resonance &  $c_{i}$ & $s_{i}$\\[0.5ex]
\hline
\hline
1 &   $\omega$&      $-1.11\pm0.09_{-0.01}^{+0.02}$ & 0.00\\[0.5ex]
2 & $K^{*-}\rho^{+}$ &     $-0.30\pm 0.05 \pm 0.01$ & $-0.03\pm 0.09_{-0.02}^{+0.01}$\\[0.5ex]
3 &  $K^{*+}\rho^{-}$ &       $ -0.41\pm0.07_{-0.01}^{+0.02}$ & $0.04\pm0.12_{-0.02}^{+0.01}$\\[0.5ex]
4 & $K^{*-}$&        $-0.79\pm0.09\pm 0.05$ &$-0.44\pm0.18\pm0.06$\\[0.5ex]
5 & $K^{*+}$ &          $-0.62\pm0.12_{-0.02}^{+0.03}$ & $0.42\pm0.20\pm0.06$ \\[0.5ex]
6 & $K^{*0}$ &         $-0.19\pm0.11\pm 0.02$ & 0.00\\[0.5ex]
7 & $\rho^{+}$ &          $-0.82\pm0.11\pm 0.03$ & $-0.11\pm0.19_{-0.03}^{+0.04}$\\[0.5ex]
8 & $\rho^{-}$ &          $-0.63\pm0.18\pm 0.03$& $0.23\pm0.41_{-0.03}^{+0.04}$ \\[0.5ex]
9 & Remainder &          $-0.69\pm0.15_{-0.12}^{+0.15}$ & 0.00\\[0.5ex]
\hline
\end{tabular}
\caption{$c_{i}$ and $s_{i}$ results in nine bins of $D$ phase space measured in CLEO-c data~\cite{Resmi}.}\label{Table:cisi}
\end{table}

\section{Data samples and event selection}

The data sample collected by the Belle~\cite{Belle1,Belle2} detector, corresponding to an integrated luminosity of 711~fb$^{-1}$, is used in this analysis. The $e^+e^-$ collisions happen at a centre-of-mass energy corresponding to the pole of the $\Upsilon(4S)$ resonance. Monte Carlo~(MC) samples are used to optimize the selection criteria, determine the efficiencies and identify various sources of background.

The decays $B^{+}\to DK^{+}$ and $B^{+} \to D\pi^{+}$ are reconstructed, in which the $D$ decays to the four-body final state of $K_{\rm S}^0\pi^+\pi^-\pi^0$. We also select $D^{*+}\to D\pi^{+}$ decays produced via the $e^+e^-\to c\bar{c}$ continuum process to measure the $K_i$ and $\overline{K_i}$ parameters. The charged particle candidates are required to come from within 0.5 cm and $\pm$ 3.0 cm of the interaction point in perpendicular and parallel directions to the $z$-axis, respectively, where the $z$-axis is defined to be opposite to the $e^+$ beam direction. These tracks are then identified as kaons or pions with the help of the particle identification system at Belle~\cite{Belle1}. The invariant masses of $K_{\rm S}^0$, $\pi^0$ and $D$ mesons are required to be within 3$\sigma$ of their nominal masses~\cite{PDG}, where $\sigma$ is the mass resolution. The background due to random combinations of pions forming a $K_{\rm S}^0$ is reduced with the help of a neural network~\cite{NB} based selection with 87\% efficiency~\cite{nisks}. The energy threshold of the photons coming from the $\pi^0$ decays are optimized according to the region in which they are detected in the electromagnetic calorimeter. Furthermore, kinematic constraints are applied to $K_{\rm S}^0$, $\pi^0$ and $D$ invariant masses and decay vertices. These constraints improve the energy and momentum resolution of the $B$ candidates and the invariant masses used to divide the $D$ phase space into bins.

The accompanying pion in the $D^{*+}\to D\pi^{+}$ decay carries a small fraction of the momentum due to the limited phase space of the decay and hence is known as a slow pion. So, while reconstructing $D^{*+}\to D\pi^{+}$ decays, it is required that the accompanying pion has at least one hit in the silicon vertex detector. The $D$ meson momentum in the laboratory frame is chosen to be between 1\textendash 4 GeV/$c$ so that the distribution matches that of the $B^+\to Dh^+ (h= K, \pi)$ sample as much as possible. The signal candidates are identified by the kinematic variables $M_D$, the invariant mass of $D$ candidate and $\Delta M$, the difference in the invariant masses of $D^*$ and $D$ candidates. The events that satisfy the criteria, $1.80< M_D < 1.95$~GeV/$c^2$ and $\Delta M < 0.15$~GeV/$c^2$ are retained. A kinematic constraint is applied so that the $D$ and $\pi$ candidates come from a common vertex position. When there is more than one candidate in an event, the one with the smallest $\chi^2$ value from the $D^*$ vertex fit is retained for further analysis. The overall selection efficiency is 3.7\%.

The $B^+$ meson candidates are reconstructed by combining a $D$ candidate with a $K^+$ or $\pi^+$ track. The $D$ meson invariant mass is selected in the range 1.835\textendash 1.890 GeV/$c^2$. We use the kinematic variables energy difference, $\Delta E$ and beam-constrained mass, $M_{\rm bc}$ to identify the signal candidates, which are  defined as
$\Delta E~=~ E_{B} - E_{\rm beam}$ and $M_{\rm bc}~=~ c^{-2}\sqrt{E_{\rm beam}^{2}-|\vec{\mathbf{p}}_{B}|^2c^{2}}$. Here $E_{B}$ and $\vec{\mathbf{p}}_{B}$ are the energy and momentum of the $B$ candidate and $E_{\rm beam}$ is the beam energy in the centre-of-mass frame. We select the candidates that satisfy the criteria $M_{\rm bc} > $~5.27 GeV/$c^2$ and $-0.13 < \Delta E < 0.30$~GeV for further analysis. In events with more than one candidate, the candidate with the smallest value of $\left(\frac{M_{\rm bc}-M_{B}^{PDG}}{\sigma_{M_{\rm bc}}}\right)^{2} + \left(\frac{M_{D}-M_{D}^{PDG}}{\sigma_{M_{D}}}\right)^{2} + \left(\frac{M_{\pi^{0}}-M_{\pi^{0}}^{PDG}}{\sigma_{M_{\pi^{0}}}}\right)^{2} $ is retained. Here, the masses $M_i^{\rm PDG}$ are those reported by the Particle Data Group in Ref.~\cite{PDG} and the resolutions $\sigma_{M_{\rm bc}},~\sigma_{M_{D}}$ and $\sigma_{M_{\pi^{0}}}$ are obtained from MC simulated samples of signal events.

The dominant background for any $B$ meson decay is due to the $e^{+}e^{-}\to q\bar{q}, q = u,~d,~s,~c$ continuum processes. Differences in the event topology between $B$ meson pairs and continuum events are used to suppress this background. The $B$ meson pairs produced from the decay of the $\Upsilon(4S)$ are almost at rest in the centre-of-mass frame, because the available energy is just above the threshold to form a $B\overline{B}$ pair. As $B$ mesons have spin zero, there is no preferred direction in space for the decay products. Thus the $B\overline{B}$ events follow a uniform spherical topology. But lighter-quark pairs are produced with large initial momentum and hence two back-to-back jets are formed in an event. These events types are separated with a NN~\cite{NB} using event shape variables and other discriminating variables including angular, vertex and flavour tag observables as input. The output of the NN is required to be greater than $-$0.6, which reduces the continuum background by 67\% and signal efficiency by 5\%. The overall selection efficiency is 4.7\% and 5.3\% for $B^+\to DK^+$ and $B^+\to D\pi^+$ modes, respectively. 

\section{Determination of $K_i$ and $\overline{K_i}$}
The $K_i$ and $\overline{K_i}$ parameters  are measured from the $D^{*+}\to D\pi^+$ sample by estimating the $D^0$ and $\overline{D^0}$ signal yields in each bin. The charge of the pion determines the flavour of the $D$ meson. The signal yield is obtained from a two-dimensional extended maximum-likelihood fit to $M_D$ and $\Delta M$ distributions independently in each bin. Appropriate probability density functions (PDF) are used to model the distributions. A quadratic correlation between $M_D$ and the width of the $\Delta M$ distribution is taken into account for the signal component. The yields along with $K_i$ and $\overline{K_i}$ values are given in Table~\ref{Table:Ki}.
\begin{table} [t] 
\centering  
\begin{tabular} {l cc cc }
\hline 
Bin no. & $N_{D^0}$ & $N_{\overline{D^0}}$ & $K_i$ & $\overline{K_i}$ \\[0.5ex]
\hline
\hline
1& $\phantom{0}$51048$\pm$282 & $\phantom{0}$50254$\pm$280 & 0.2229$\pm$0.0008 & 0.2249$\pm$0.0008\\[0.5ex]
2& 137245$\pm$535 & $\phantom{0}$58222$\pm$382 & 0.4410$\pm$0.0009 & 0.1871$\pm$0.0007\\[0.5ex]
3& $\phantom{0}$31027$\pm$297 & 105147$\pm$476 & 0.0954$\pm$0.0005 & 0.3481$\pm$0.0009\\[0.5ex]
4& $\phantom{0}$24203$\pm$280 & $\phantom{0}$16718$\pm$246 & 0.0726$\pm$0.0005 & 0.0478$\pm$0.0004\\[0.5ex]
5& $\phantom{0}$13517$\pm$220 & $\phantom{0}$20023$\pm$255 & 0.0371$\pm$0.0003 & 0.0611$\pm$0.0004\\[0.5ex]
6& $\phantom{0}$21278$\pm$269 & $\phantom{0}$20721$\pm$267 & 0.0672$\pm$0.0005 & 0.0679$\pm$0.0005\\[0.5ex]
7& $\phantom{0}$15784$\pm$221 & $\phantom{0}$13839$\pm$209 & 0.0403$\pm$0.0004 & 0.0394$\pm$0.0004\\[0.5ex]
8& $\phantom{00}$6270$\pm$148 & $\phantom{00}$7744$\pm$164 & 0.0165$\pm$0.0002 & 0.0183$\pm$0.0002\\[0.5ex]
9& $\phantom{00}$6849$\pm$193 & $\phantom{00}$6698$\pm$192 & 0.0070$\pm$0.0002 & 0.0054$\pm$0.0001\\[0.5ex]

\hline
\end{tabular}
\caption{$D^0$ and $\overline{D^0}$ yield in each bin of $D$ phase space along with $K_i$ and $\overline{K_i}$ values measured in $D^{*}$ tagged data sample.}\label{Table:Ki}
\end{table}

\section{Estimation of $x_{\pm}$ and $y_{\pm}$}
The signal yield in each $D$ phase space bin of $B^+\to D\pi^+$ and $B^+\to DK^+$ decays is determined from a two-dimensional extended maximum-likelihood fit to $\Delta E$ and neural network output ($C_{\rm NN}$). The latter is transformed as
\begin{equation}
C_{\rm NN}' = \log\left( \frac{C_{\rm NN} - C_{\rm NN, low}}{C_{\rm NN, high} - C_{\rm NN}} \right),
\end{equation}
where $C_{\rm NN, low}$ = $-0.6$ and $C_{\rm NN, high} = 0.9985$ are the minimum and maximum values of $C_{\rm NN}$ in the sample, respectively. There are three types of background events: continuum background, combinatorial $B\overline{B}$ background due to final state particles from both the $B$ mesons and cross-feed peaking background due to the misidentification of a pion as a kaon and {\it vice versa}.
\begin{figure}[ht!]
\centering
\includegraphics[width=0.8\columnwidth]{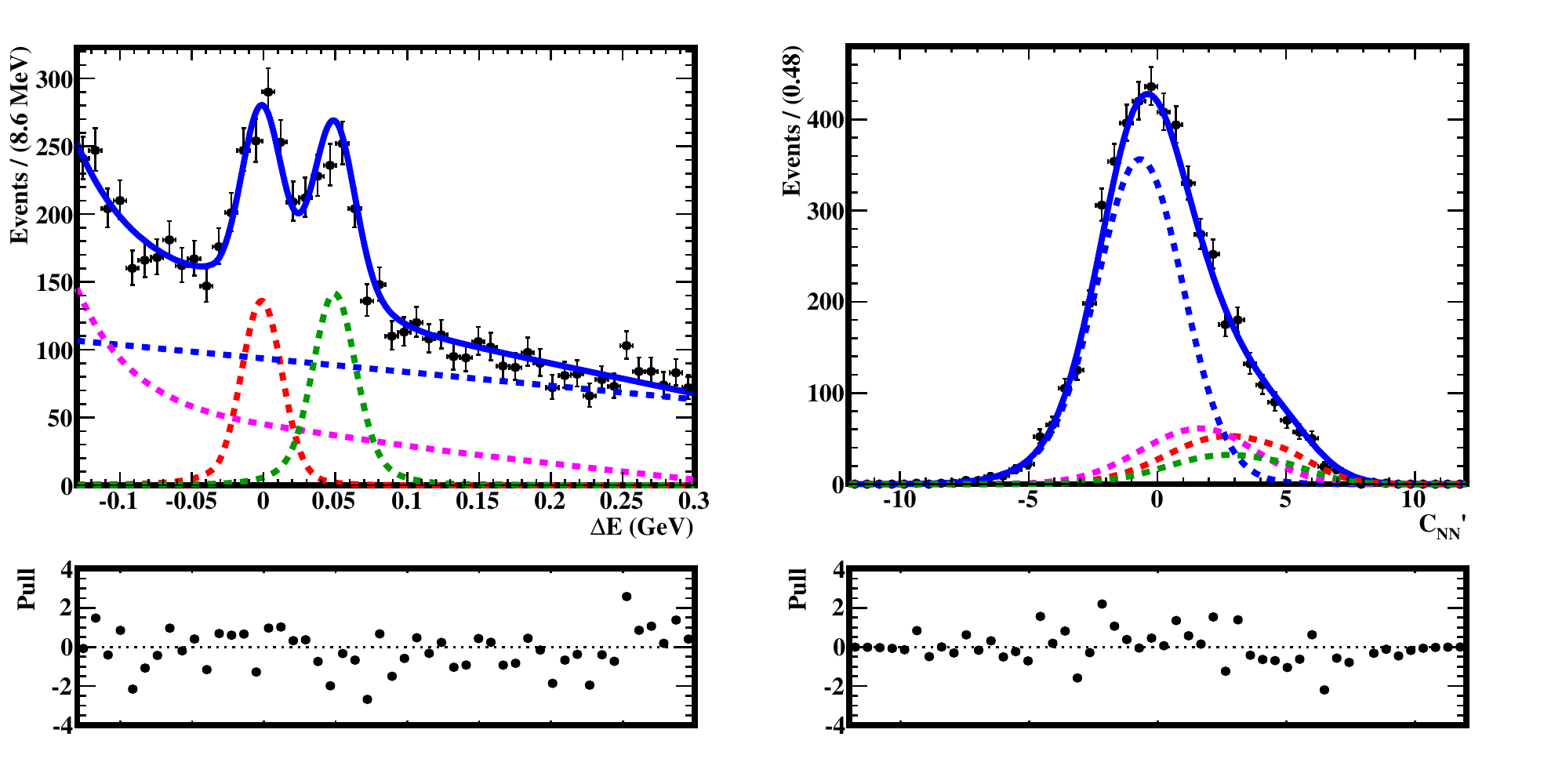}
\caption{Signal-enhanced fit projections of $\Delta E$ (left) and $C_{\rm NN}'$ (right) in data for $B^{\pm}\to DK^{\pm}$ decays. The black points with error bar are the data and the solid blue curve is the total fit. The dotted red, blue, magenta and green curves represent the signal, continuum, random $B\overline{B}$ backgrounds and cross-feed peaking background components, respectively. The pull between the data and the fit are shown for both the projections. }\label{Fig:DK}
\end{figure}
The fit projections in data sample for $B^+\to DK^+$ decays are shown in Fig.\ref{Fig:DK}. These are signal-enhanced projections with events in the signal region of the other variable, where the signal regions are defined as $|\Delta E| < 0.05$ GeV and $C_{\rm NN}'>0$. The fit model is verified in MC samples and pseudo-experiments are performed to check for any possible bias.

\begin{figure}[ht!]
\centering
\begin{tabular}{cc}
\includegraphics[width=0.4\columnwidth]{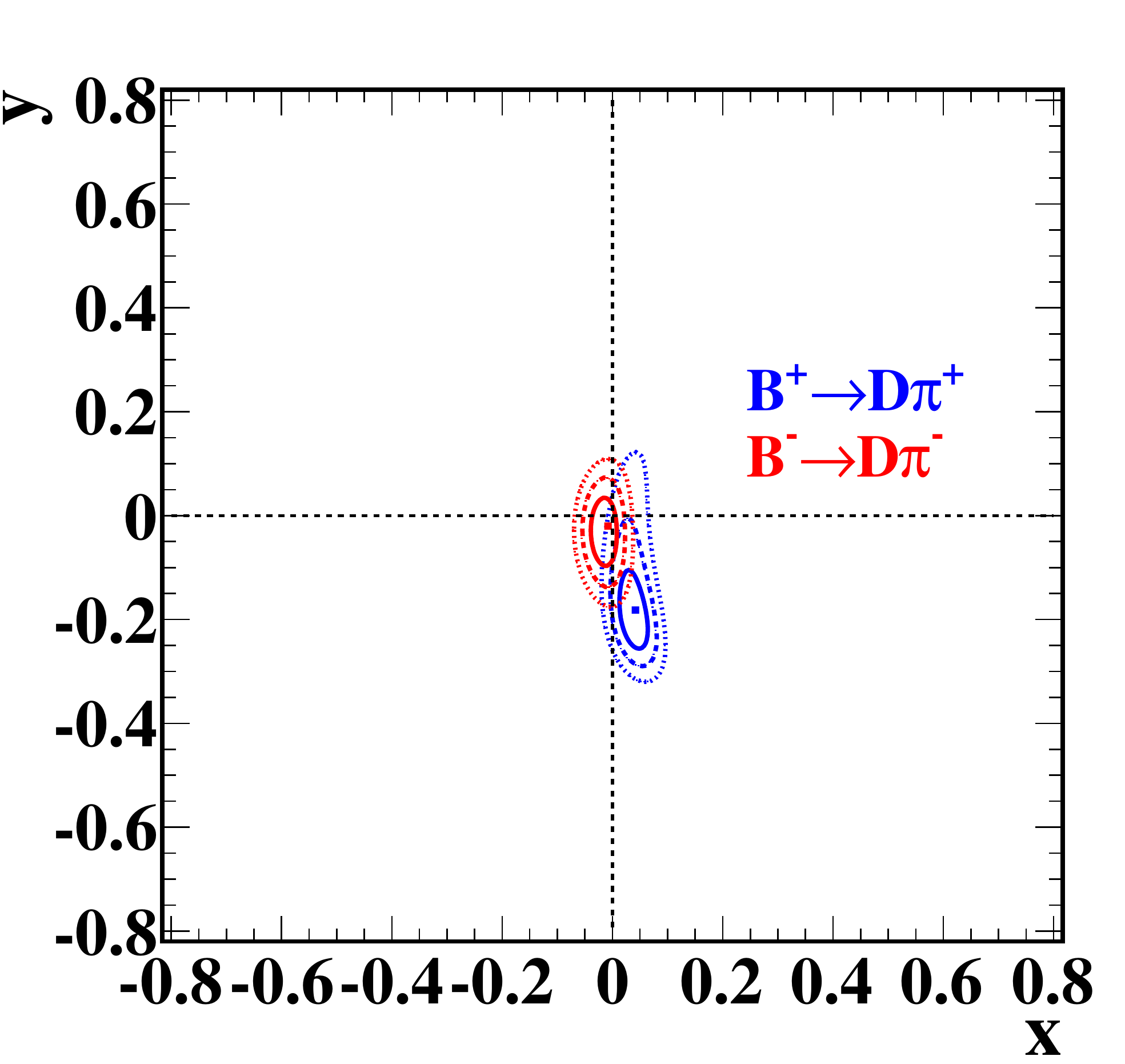}&
\includegraphics[width=0.4\columnwidth]{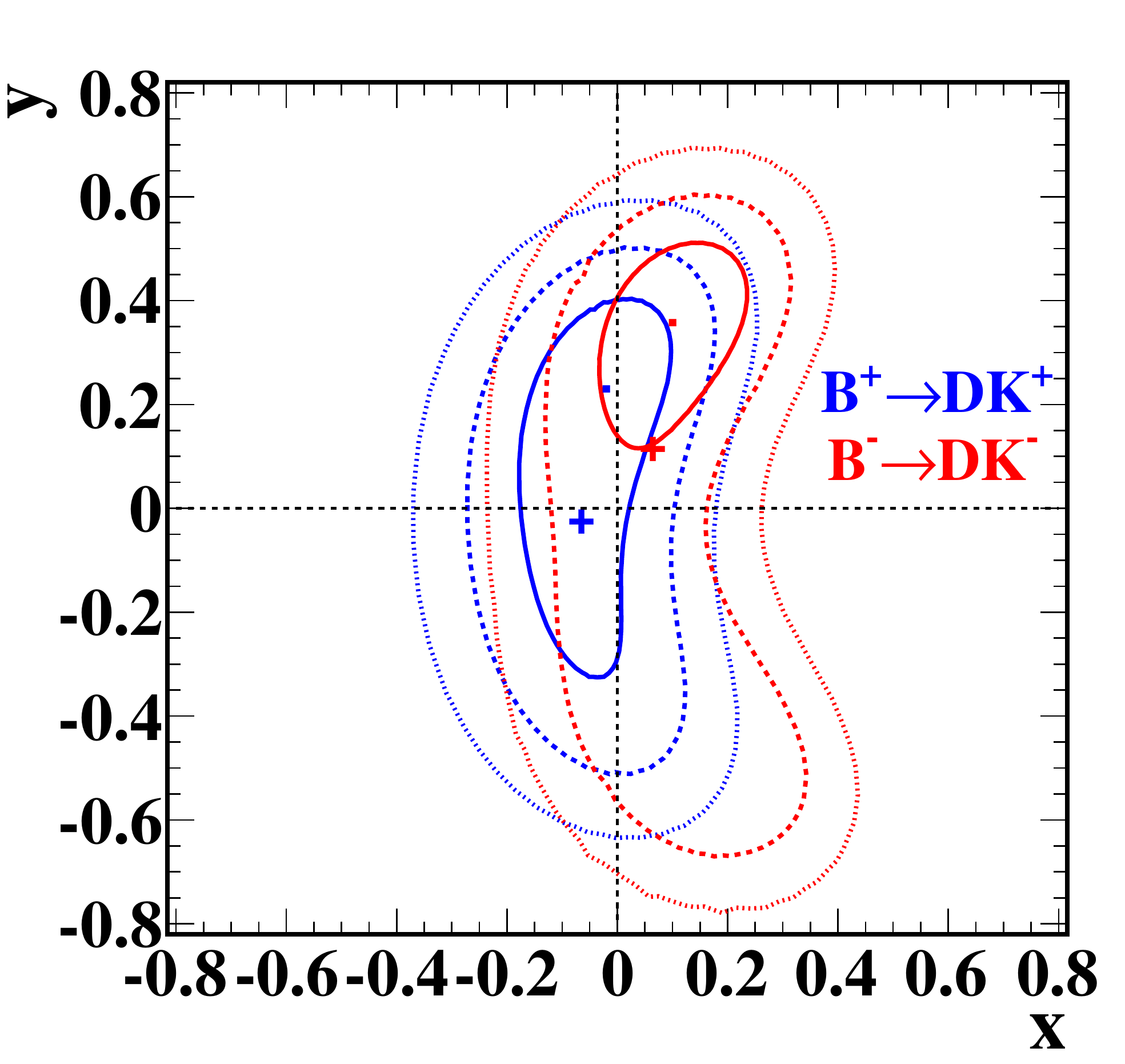}\\
\end{tabular}
\caption{One (solid line), two (dashed line), and three (dotted line) standard deviation  likelihood contours for the $(x_{\pm},y_{\pm})$ parameters for $B^{\pm}\to D\pi^{\pm}$ (left) and  $B^{\pm}\to DK^{\pm}$ (right) decays in data. The point marks the best fit value and the cross marks the expected value from the world average values of $\phi_3$, $r_{B}^{DK}$, and $\delta_{B}^{DK}$ \cite{HFLAV}.}
\label{Fig:contour}
\end{figure}

The fit is performed simultaneously to the nine bins and the $\phi_3$-sensitive parameters $x_{\pm}$ and $y_{\pm}$ are determined directly from the fit by expressing the signal yield as in Eq.~\ref{Eq:GGSZ}. The measured values of $K_i$ and $\overline{K_i}$ in Table~\ref{Table:Ki} along with the $c_i$ and $s_i$ measurements reported in Ref.~\cite{Resmi} are used as input parameters. Efficiency corrections are applied and the effect of migration of events between the bins due to finite momentum resolution is also taken into account via a migration matrix. The results obtained are given in Table~\ref{Table:xy_data}.  The statistical likelihood contours are shown in Fig.~\ref{Fig:contour}.
\begin{table} [t] 
\centering  
\begin{tabular} {l c c }
\hline 
\hline
& $B^{\pm}\to D\pi^{\pm}$ & $B^{\pm}\to DK^{\pm}$ \\[1ex]
\hline
\\[-1em]
$x_{+}$ & 0.039 $\pm$ 0.024~$^{+0.018~+0.014}_{-0.013~-0.012}$ & $-$0.030 $\pm$ 0.121~$^{+0.017~+0.019}_{-0.018~-0.018}$     \\[1ex]

$y_{+}$ & $-$0.196~$^{+0.080~+0.038~+0.032}_{-0.059~-0.034~-0.030}$ & 0.220~$^{+0.182}_{-0.541}$  $\pm$ 0.032~$^{+0.072}_{-0.071}$   \\[1ex]

$x_{-}$ & $-$0.014 $\pm$0.021~$^{+0.018~+0.019}_{-0.010~-0.010}$ & 0.095 $\pm$ 0.121~$^{+0.017~+0.023}_{-0.016~-0.025}$ \\[1ex]

$y_{-}$  & $-$0.033 $\pm$~0.059$^{+0.018~+0.019}_{-0.019~-0.010}$ & 0.354~$^{+0.144~+0.015~+0.032}_{-0.197~-0.021~-0.049}$ \\[1ex]
\hline
\hline
\end{tabular}
\caption{$x_{\pm}$and $y_{\pm}$ parameters from a combined fit to $B^{\pm}\to D\pi^{\pm}$ and $B^{\pm}\to DK^{\pm}$ data samples. The first uncertainty is statistical, the second is systematic, and the third is due to the uncertainty on the $c_i$, $s_i$ measurements. }\label{Table:xy_data}
\end{table}

The dominant source of systematic uncertainty is due to the uncertainty on $c_i$ and $s_i$ input values. The next largest source of systematic uncertainty is the statistics of the signal MC sample used to calculate the efficiency and migration matrix. If the signal MC statistics is further increased, the data-MC resolution difference will be worse. As this measurement is statistically dominated, any small improvements in systematic uncertainty will have negligible impact.

\section{Determination of $\boldsymbol{\phi_3,~r_B}$ and $\boldsymbol{\delta_B}$}
\begin{figure}[ht!]
\centering
\begin{tabular}{cc}
\includegraphics[width=0.4\columnwidth]{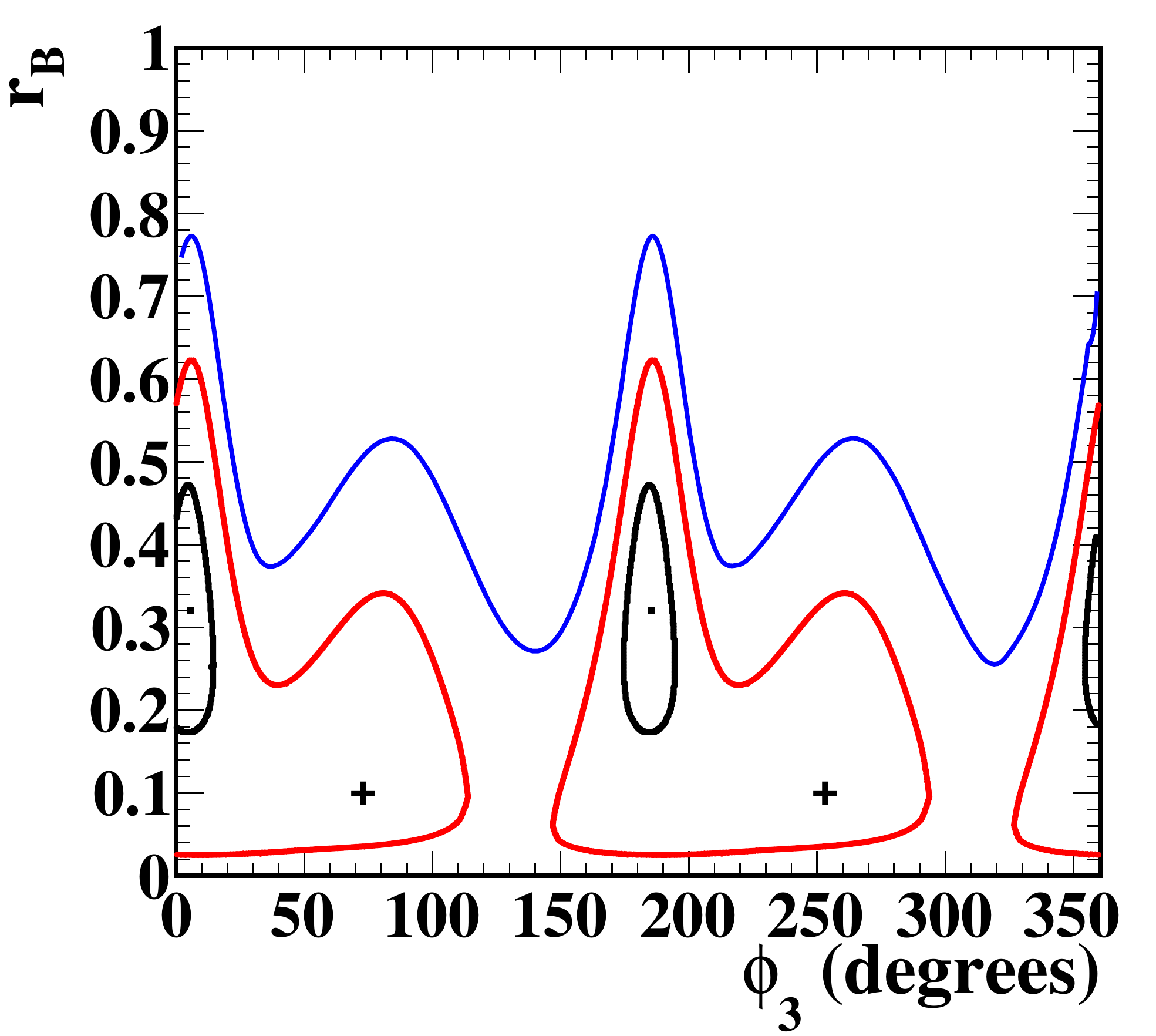}&
\includegraphics[width=0.4\columnwidth]{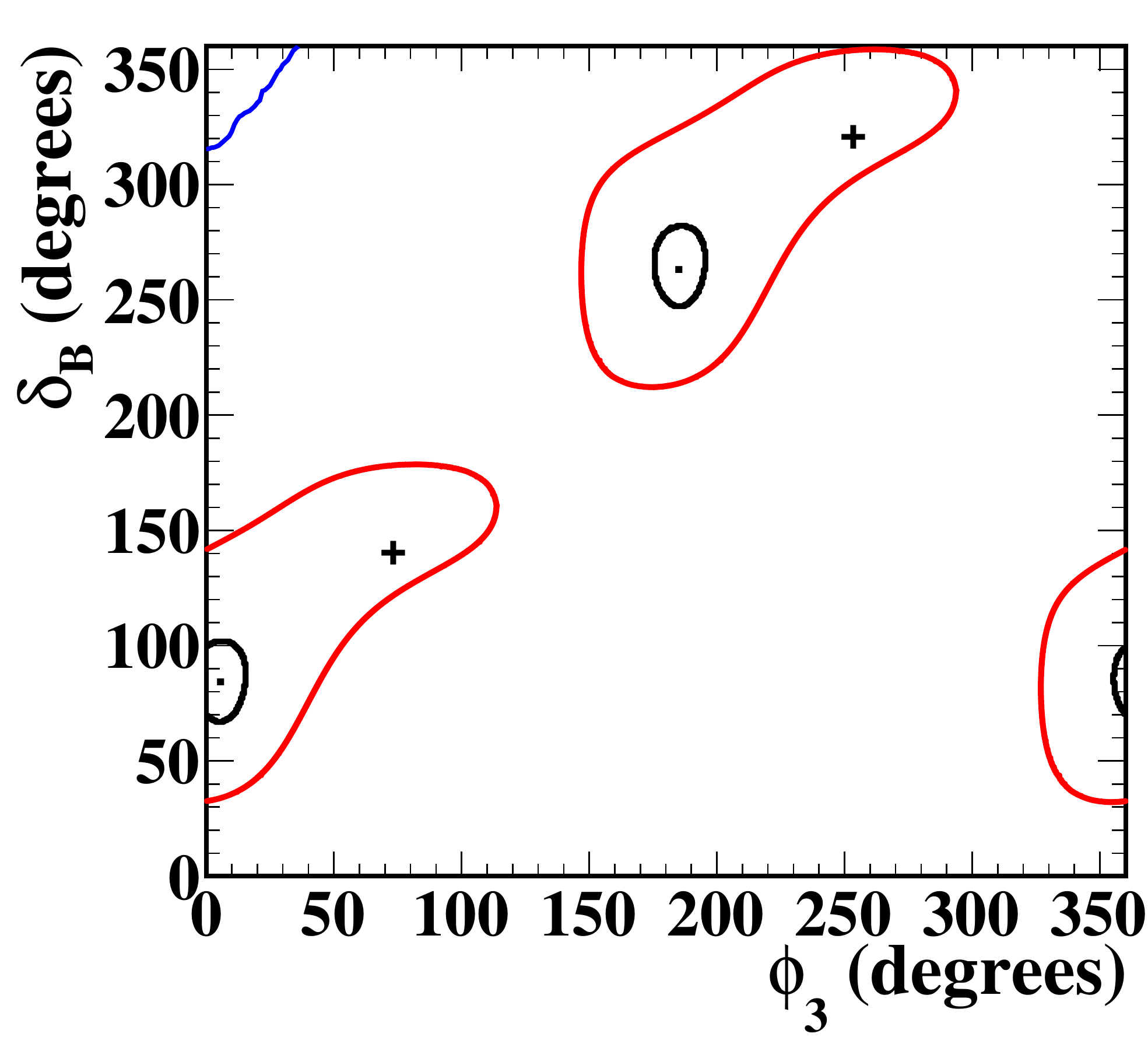}\\
\end{tabular}
\caption{Projection of the statistical confidence intervals in the $\phi_3 - r_B$ (left) and $\phi_3 - \delta_B$ (right)  planes. The black, red, and blue contours represent the one, two, and three standard deviation regions, respectively. The crosses show the positions of the world-average values~\cite{HFLAV}.}
\label{Fig:contours}
\end{figure}

We use the frequentist treatment, which includes the Feldman-Cousins ordering~\cite{FC}, to obtain the physical parameters $(\phi_3, r_{B}, \delta_{B})$ from the measured parameters $(x_{+}, y_{+}, x_{-}, y_{-})$ in $B^{\pm}\to DK^{\pm}$ sample; this is the same procedure as used in Ref.~\cite{Belle-GGSZ}.  We obtain the parameters $(\phi_3, r_{B}, \delta_{B})$ as given in Table~\ref{Table:phi3}. Figure~\ref{Fig:contours} shows the statistical confidence level contours representing the one, two, and three standard deviation in $(\phi_3, r_B)$ and $(\phi_3, \delta_B)$ planes. 
\begin{table} [ht!] 
\centering  
\begin{tabular} {l c c c  c  }
\hline 
\hline
Parameter & Results & 2$\sigma$ interval\\[1ex]
\hline
\\[-1em]
$\phi_3$ ($^{\circ}$) & $ 5.7~^{+10.2}_{-8.8}~ \pm~ 3.5~ \pm~ 5.7$  & ($-$29.7 , 109.5)\\[0.5ex]
$\delta_B$ ($^{\circ}$) & $83.4~^{+18.3}_{-16.6}~ \pm~ 3.1~ \pm~ 4.0 $ & (35.7 , 175.0)\\[0.5ex]
$r_B$ & $0.323~ \pm~ 0.147~ \pm~ 0.023~ \pm~ 0.051$& (0.031 , 0.616)\\[0.5ex]
\hline
\hline
\end{tabular}
\caption{$(\phi_3, \delta_B , r_B)$ obtained from the $B^{\pm}\to DK^{\pm}$ data sample. The first uncertainty is statistical, second is systematic and, the third one is due to the uncertainty on $c_i$, $s_i$ measurements.}\label{Table:phi3}
\end{table}

We combine the results presented here with the model-independent measurements from the decays  $B^{+}\to~D(K_{\rm S}^0\pi^+\pi^-)K^{+}$~\cite{Belle-GGSZ} and $B^0\to~D^0(K_{\rm S}^0 \pi^+ \pi^-) K^{*0}$~\cite{Belle-GGSZ2}, which use the full dataset collected by the Belle detector. Without our measurement, the combination leads to $\phi_3~=~(78^{+14}_{-15})^{\circ}$. Including our measurement, the combination gives $\phi_3~=~(74^{+13}_{-14})^{\circ}$. 

\section{Summary}
The precise measurement of the CKM angle $\phi_3$ is important to further test the $CP$ violation mechanism in the SM. We find that $D\to K_{\rm S}^0\pi^+\pi^-\pi^0$ is a promising candidate to add to the $D$ final states that determine $\phi_3$, due to its larger branching fraction and resonance substructures. We have measured $\phi_3$ from $B^{\pm} \to D( K_{\rm S}^0\pi^+\pi^-\pi^0)K^{\pm}$ decays for the first time. This decay mode is expected to provide a sensitivity of 4.4$^{\circ}$ with 50~ab$^{-1}$ data from Belle~II. Further improvements are possible once an amplitude model for $D^{0}\to K_{\rm S}^0\pi^{+}\pi^{-}\pi^{0}$ is available to guide the binning of the phase space such that maximum sensitivity to $\phi_3$ is obtained.  A more precise measurement of $c_i$ and $s_i$ parameters could be performed with a larger sample of $e^{+}e^{-}\to \psi(3770)$ data that has been collected by BESIII, thus reducing the systematic uncertainty.

\end{document}